# Conical Intersections Shed Light on Hot Carrier Cooling in Quantum Dots

Caitlin V. Hetherington[1], Nila Mohan T. M.[2], Shanu A. Shameem[2], Warren F. Beck[2], Benjamin G. Levine[1*]

[1] Institute for Advanced Computational Science and Department of Chemistry, Stony Brook University, Stony Brook, New York 11733 U.S.A.

[2] Department of Chemistry, Michigan State University, East Lansing, Michigan 48824 U.S.A.

* email: ben.levine@stonybrook.edu

## Abstract

Experimental observations of vibronic coherences in electronically excited colloidal semiconductor nanocrystals offer a window into the ultrafast dynamics of hot carrier cooling. In previous work, we showed that, in amine-passivated quantum dots (QDs), these coherences arise during relaxation through a cascade of conical intersections between electronically excited states. Here, we demonstrate the generality of this framework by application to QDs with surface-bound carboxylate ligands. A model involving a similar cascade of conical intersections accurately reproduces the frequencies of vibronic coherences observed with broadband multidimensional spectroscopy. The impact of ligands on the relaxation dynamics is attributed to two distinct mechanisms involving either electronic or vibrational coupling between the core and ligands. Compared to the amine-passivated QDs studied previously, the electronic coupling mechanism is less prominent in carboxylate-passivated QDs. Furthermore, comparison of acetate and formate ligands reveals that truncating the ligand alkyl chains alters the relaxation behavior predicted by the model.

## 1. Introduction

Owing to their size-, composition-, and surface-dependent properties, quantum dots (QDs) are promising candidates for light-harvesting applications.[1-5] However, the maximum efficiency of a single-junction solar cell is fundamentally limited by the Shockley–Queisser limit.[6] This limit arises in large part from hot carrier cooling—a process in which excess energy above the band gap is dissipated via nonradiative relaxation in the conduction band following photoexcitation.[7, 8] To surpass this efficiency barrier, hot carriers must be extracted before they cool. This might be achieved either by slowing the relaxation process and enhancing the predicted phonon bottleneck,[9-18] or by promoting charge transfer to surface-bound acceptors on a timescale faster than carrier relaxation.[19-21] Designing QDs with such tailored behavior requires a detailed understanding of the mechanisms governing hot carrier dynamics.

It is well-established that the Auger mechanism plays a central role in hot carrier cooling.[14, 22-31] In addition, it has been shown that surface ligands significantly influence the cooling rate,[18, 32-37] suggesting that tuning the surface composition offers a viable strategy for control. In parallel, computational and spectroscopic studies have revealed that the QD–ligand interface supports photoinduced ligand-mediated processes,[38-44] including ligand detachment and reattachment in the excited state, which can in turn drive energy transfer.[45]

Moreover, recent advances in nonlinear optical spectroscopy[46] have enabled direct investigation of the mechanisms underlying the initial ultrafast steps of hot carrier cooling. In particular, multidimensional electronic spectroscopy—including two-dimensional[47] (2DES) and three-dimensional (3DES) spectroscopies[48]—has enabled a detailed picture of the carrier dynamics in QDs.[49-65] Using broadband 2DES and 3DES, Tilluck et al. demonstrated that subpicosecond electronic relaxation in amine-passivated CdSe QDs is promoted by coherences of mixed vibrational and electronic character, involving strong coupling between electronic excitations and ligand vibrations.[63]

Building on these findings, our recent work explored the origin of these coherences using a novel theoretical framework inspired by molecular photochemistry.[66] Our approach is based on the hypothesis that conical intersections (CIs) mediate relaxation between closely spaced exciton states in the QD. CIs—points of degeneracy between two electronic state potential energy surfaces (PESs)—are well-known to facilitate ultrafast nonadiabatic transitions between electronic states.[67-69] In amine-passivated dots, a model assuming relaxation through a cascade of electronic states accurately reproduces the frequencies of the experimentally-observed vibronic coherences. Thus relaxation through conical intersections provides a viable explanation for the experimentally observed dynamics.

Furthermore, our analysis identified two distinct pathways by which surface ligands can influence the rate of cooling. In the *vibrational mechanism*, electronic excitation of the QD core induces vibrational excitation of the core, which in turn drives vibrational excitation of the ligands via harmonic coupling between modes of similar frequency. In the *electronic mechanism*, delocalization of the initial electronic excitation onto the ligands leads directly to vibrational motion in the ligands. Both mechanisms operate in amine-passivated QDs: the vibrational mechanism dominates at low frequencies, where coupling to CdSe phonon modes is possible, while the electronic mechanism dominates at higher frequencies, where core modes are absent.

In this work, we investigate whether our CI approach is general by extending it to QDs passivated with chemically different ligands on the surface. Forthcoming work by the Beck group reveals the presence of coherences in the first ultrafast steps of hot carrier cooling upon photoexcitation of oleate-passivated QDs. We apply our CI approach to two small model carboxylate-passivated CdSe QDs to analyze the vibrational modes involved in hot carrier cooling and the mechanisms by which they are driven. Furthermore, by comparison of acetate- and formate-passivated QDs, we investigate the role that termination of the ligands plays in our model. In section 2, we describe the details of our computational study. In section 3, we present results and discussion, and finally, in section 4, we draw conclusions.

## 2. Computational Details

In this work, we estimate the energy deposited into each vibrational mode of carboxylate-passivated QDs upon relaxation through a cascade of conical intersections, using a procedure developed in our previous work.[66] Our approach is based on the key assumption that the QD relaxes via a piecewise straight-line path from the Franck–Condon (FC) point, through a series of minimum-energy conical intersections (MECIs) connecting each pair of states (X4→X3→X2→X1), and ultimately to the minimum-energy geometry of the lowest excited state (X1). Note that this path does not correspond directly to the experimental data, in which exciation to X3 is considered. However, the paths are likely similar as the only difference is the inclusion of the X4-X3 MECI. Projecting our hypothetical path onto the vibrational normal modes yields a spectral density. We identify the modes with the largest contributions as those most likely to be excited during the relaxation process and therefore most likely to appear as coherent oscillations in the 2DES.

To simplify the analysis, we assume that the diabatic excited states share the same normal modes and frequencies as the ground state (i.e., no Duschinsky rotation), and that each MECI lies at the minimum of the upper diabatic state. While this latter assumption is certainly not quantitatively accurate, it affects only the amplitudes of peaks in the spectral density, not their positions. As discussed below, these peak heights do not correspond directly to any spectroscopic observable and therefore their precise values are of limited interpretive consequence.

In line with the experiment, we assume an initial FC excitation to the fourth excited state (X4) at the FC point (ground state minimum structure). Given an optimized FC point, $R_{FC}$, optimized MECIs between each pair of excited states from X4 down to X1, ($R_{CI_{X4-X3}}$, $R_{CI_{X3-X2}}$...) and the optimized first excited state minimum structure, $R_{Min_{X1}}$, we compute the spectral density as follows. We first calculate the displacement vector, $\Delta R_{CI_{X4-X3}-FC}$, connecting FC geometry and the geometry of the first CI, CI$_{X4-X3}$,

$$\Delta R_{CI_{X4-X3}-FC} = R_{CI_{X4-X3}} - R_{FC}. \qquad 1$$

We similarly compute displacement vectors for each segment of the relaxation pathway: CI$_{X4-X3}$ to CI$_{X3-X2}$, CI$_{X3-X2}$ to CI$_{X2-X1}$, and CI$_{X2-X1}$ to Min$_{X1}$. Note that all displacement vectors, $\Delta R_J$ (where $J$ indexes the corresponding segment of the relaxation pathway) are represented in mass-weighted Cartesian coordinates.

Then each vector, $\Delta R_J$, is projected onto the ground-state mass-weighted normal modes to yield its representation in normal coordinates, $\Delta q_J$. Finally, the energy, $\Delta E_i$, associated with each normal mode, $i$, of vibrational frequency $\omega_i$, is calculated by summing over all segments according to,

$$\Delta E_i = \tfrac{1}{2}\omega_i^2 \sum_J \Delta q_{J,i}^2. \qquad 2$$

The spectral density is $\Delta E_i$ as a function of $\omega_i$, plotted either as a stick spectrum or as its Gaussian convolution. An illustration of our approach is shown in figure S1.

This approach takes into account the complexity of the PESs in the vicinity of CIs, recognizing that passage through a CI breaks the system's symmetry. As a result, vibrational motion is driven by more than just the Franck–Condon impulse (Figure S2). In addition, our approach is not limited to single-phonon emission, an approximation that has recently been shown to fail to accurately describe carrier cooling.[70] Furthermore, because CIs are topologically protected, they are not typically eliminated by small changes in system size, shape, or environment, making the results of this approach extrapolatable to larger experimental systems. However, this approach has limitations—chiefly the neglect of anharmonicity and the assumption that each CI lies at the minimum of the diabatic potential.

In addition to evaluating the full reaction coordinate described above (FC→$CI_{X4-X3}$→$CI_{X3-X2}$→$CI_{X2-X1}$ →$Min_{X1}$), we also evaluate the direct path from GS to $Min_{X1}$, which bypasses the CIs. The goal is to determine whether invoking CIs is necessary to explain the experimentally observed coherences. To this end, we apply the same procedure, replacing the piecewise linear path with a single segment from GS to $Min_{X1}$.

Calculations were performed on the Seawulf GPU cluster at Stony Brook University and the Expanse GPU cluster at the San Diego Supercomputer Center, through the NSF ACCESS program.[71] Ground state structures were optimized with density functional theory (DFT) at the CAM-B3LYP[72]/LANL2DZ[73] level of theory. Excited states were calculated with linear-response time-dependent density functional theory[74] (TDDFT) with the same functional and basis set. All calculations were performed with the TeraChem[75-77] software package (version v1.9-2023.03-dev). Minimal energy conical intersections between electronic excited states were optimized with the CIOpt[78] software package. We note that conical intersections between pairs of electronic excited states are well-treated by TDDFT, despite well-known issues with intersections involving the DFT ground state.[79] The root mean square distances (RMSDs) between pairs of geometries were obtained using the VMD visualization package.[80] Ground state normal modes were computed with DFT at the PBE[81]/LANL2DZ level of theory, using the restart-friendly Hessian implementation in the PySpawn software package.[82]

## 3. Results and Discussion

To investigate the initial ultrafast steps in hot carrier cooling following photoexcitation of carboxylate-passivated CdSe QDs, we first analyzed the potential energy surfaces (PESs) of two model systems. Both models contain a $Cd_{33}Se_{33}$ core,[34] but they differ in the nature of the carboxylate ligands on their surfaces. The first QD, $Cd_{33}Se_{33}(CO_2H)_9H_9$, is passivated with nine formate ligands (optimized geometry shown at the top of Figure 1a). The second QD, $Cd_{33}Se_{33}(CO_2Me)_9H_9$, features nine acetate ligands (optimized geometry shown at the top of Figure 1b). In both cases, nine protons were added to surface selenium atoms to neutralize the system. Each QD has a diameter of approximately 1.3 nm.

## 3.1. Relaxation via Cascade of CIs

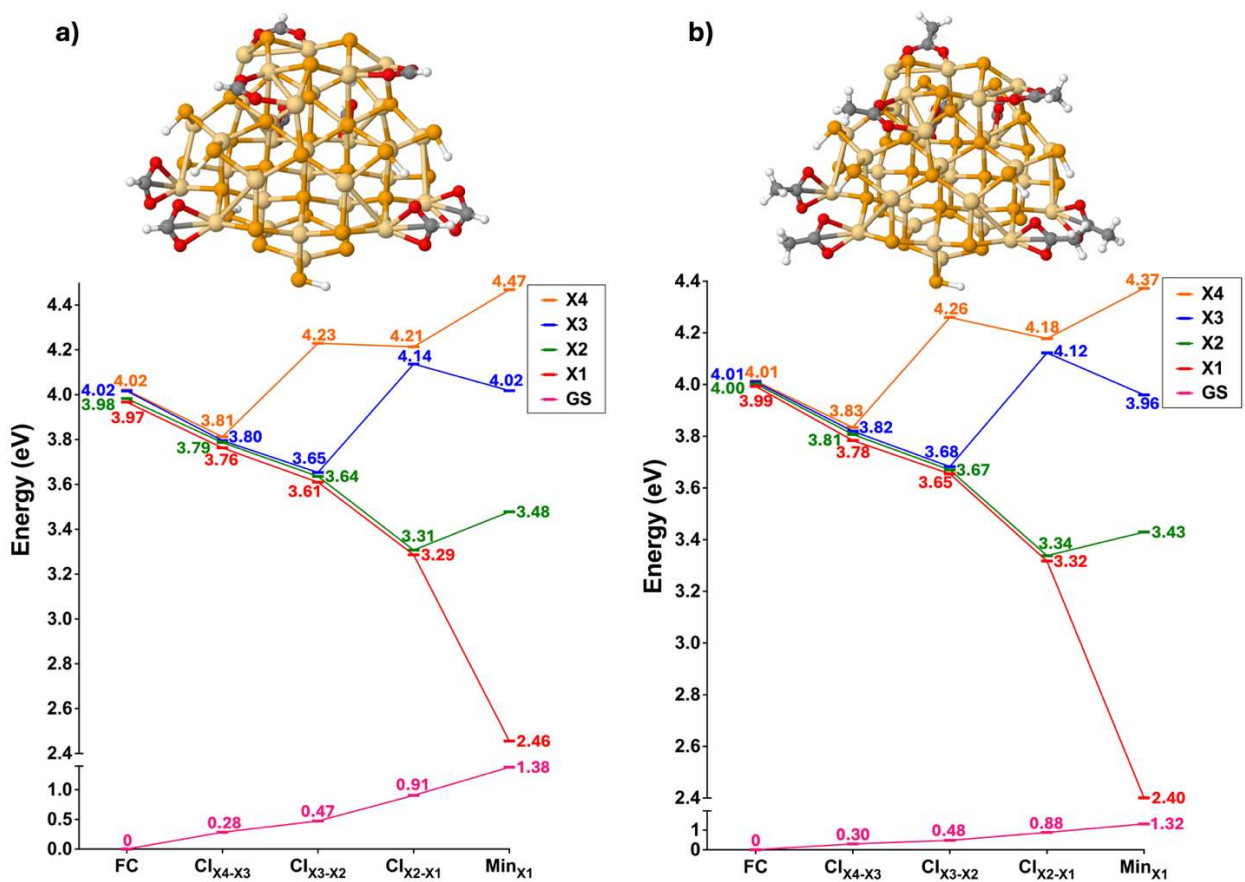

**Figure 1.** Top: Optimized geometry of a) formate-passivated and b) acetate-passivated model QDs (beige: Cd, orange: Se, red: O, gray: C, white: H). Bottom: Energies (in eV) of the ground

and first four exciton states at the Franck−Condon (FC), conical intersection ($CI_{X4-X3}$, $CI_{X3-X2}$ and $CI_{X2-X1}$), and X1 minimum ($Min_{X1}$) geometries of a) formate-passivated and b) acetate-passivated model QDs. Energies are shifted such that the ground state energy at the FC point is zero.

We computed the TDDFT energies of the first four exciton states (X1, X2, X3, and X4) at the Franck–Condon (FC) geometry, at each conical intersection ($CI_{X4–X3}$, $CI_{X3–X2}$, and $CI_{X2–X1}$), and at the minimum energy geometry on the X1 surface ($Min_{X1}$). Figure 1 illustrates the cascade of conical intersections that defines one possible reaction coordinate for hot carrier cooling. In both systems, we observe three conical intersections of decreasing energy that connect the initial excited state (X4) to X3, X2, and finally X1. Relaxation on X1 leads to the minimum-energy structure ($Min_{X1}$). We previously observed a similar cascade in an amine-passivated QD model,[64] suggesting that this relaxation mechanism may be general across various classes of QDs, regardless of the nature of the surface ligands.

To further characterize the geometrical changes during this relaxation, we calculated root-mean-square deviations (RMSDs) between the FC geometry and each CI, as well as between the FC geometry and $Min_{X1}$, excluding hydrogen atoms from the comparison (Figure S3). In both systems, $Min_{X1}$ exhibits the largest deviation from the FC point, consistent with gradual vibrational displacements as the QD relaxes through the CI cascade from FC to $Min_{X1}$.

## 3.2. Assignment of Vibronic Coherences

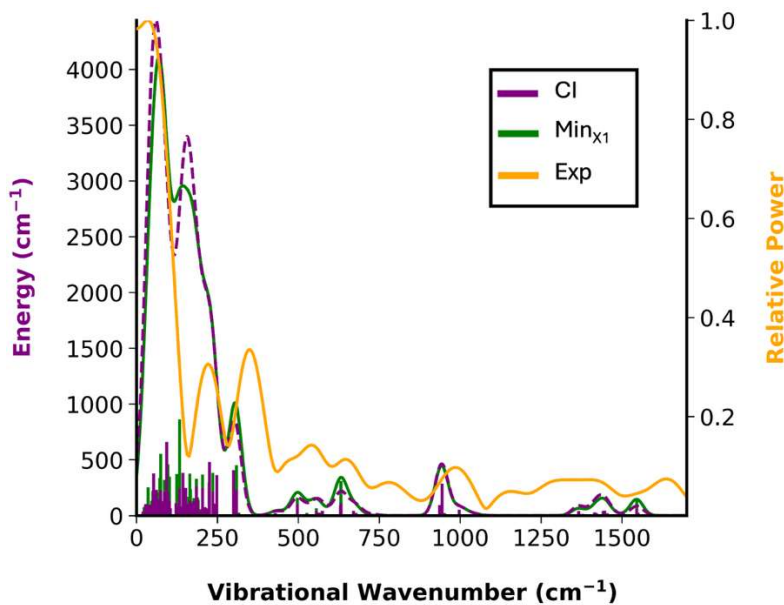

**Figure 2.** Spectral densities (in cm$^{-1}$) of the acetate-passivated QD. Purple lines and sticks correspond to the full relaxation path from FC to the minimum of X1 (Min$_{X1}$), passing through CI$_{X4-X3}$, CI$_{X3-X2}$, and CI$_{X2-X1}$. The green line and sticks correspond to a direct linear relaxation pathway from FC to Min$_{X1}$. For comparison, the Fourier transform amplitude spectrum of the X3-PL cross peak, described in the text, is shown in orange.

Having established the existence of cascades of CIs through which relaxation may occur in both models, we now analyze the vibrational modes involved in each system by computing the spectral density, to assign vibronic coherences observed in multidimensional electronic spectroscopic measurements. We first discuss the spectral density of the acetate-passivated model, tracing the relaxation path from FC to Min$_{X1}$ via CI$_{X4-X3}$, CI$_{X3-X2}$, and CI$_{X2-X1}$ (purple sticks in Figure 2). The purple dashed line in Figure 2 corresponds to the Gaussian-broadened spectral density. For comparison, experimental data obtained from multidimensional electronic spectroscopy measurements of oleate-passivated QDs by the Beck group are shown as orange lines. The spectroscopic data here focuses on a cross peak in the spectrum with an excitation energy near X3 (17762 cm$^{-1}$) and detection energy near the photoluminescence (PL) energy (15150 cm$^{-1}$). This spectrum was obtained by Fourier transformation of the oscillatory part of the 3DES at this cross peak over delay times ranging from 5 to 500 fs. The resulting peaks correspond to experimentally observed coherences created during relaxation from X3 to the band edge. It is important to note that our simulated spectral density does not account for transition dipole matrix elements, so the peak heights are not expected to match the experimental observables directly. In this comparison, we focus on the positions of the peaks rather than their intensities.

The theoretical spectrum (purple) exhibits two main clusters of peaks in the low-frequency region at 93 cm$^{-1}$ and 225 cm$^{-1}$, which align well with the two low-frequency peaks observed experimentally at 33 cm$^{-1}$ and 222 cm$^{-1}$. In the higher-frequency region (above 1250 cm$^{-1}$), small peaks appear in the calculated spectrum at 1446 cm$^{-1}$ and 1545 cm$^{-1}$, which correspond to experimental peaks at 1394 cm$^{-1}$ and 1635 cm$^{-1}$. Additionally, several intermediate-frequency features in the theoretical spectrum—at 300 cm$^{-1}$, 497 cm$^{-1}$, 631 cm$^{-1}$, and 944 cm$^{-1}$—match vibrational modes in the experimental data at 352 cm$^{-1}$, 541 cm$^{-1}$, 645 cm$^{-1}$, and 984 cm$^{-1}$. This agreement suggests that the observed coherences may originate from relaxation via a cascade of CIs, as suggested by theory.

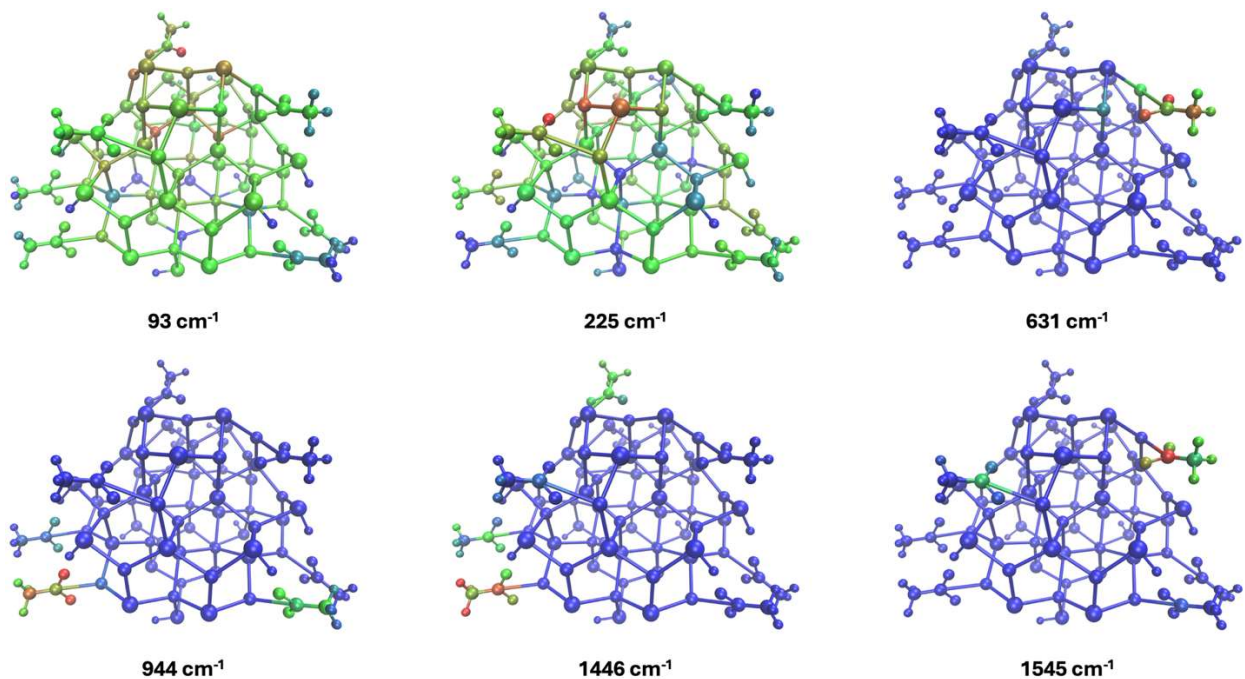

**Figure 3.** Several vibrational normal modes of the acetate-terminated QD involved in relaxation through the cascade of conical intersections ($CI_{X4-X3}$, $CI_{X3-X2}$, and $CI_{X2-X1}$) to the minimum of X1 ($Min_{X1}$). Atoms are colored according to the magnitude of the corresponding mass-weighted normal mode elements, with red indicating larger elements and blue indicating smaller elements.

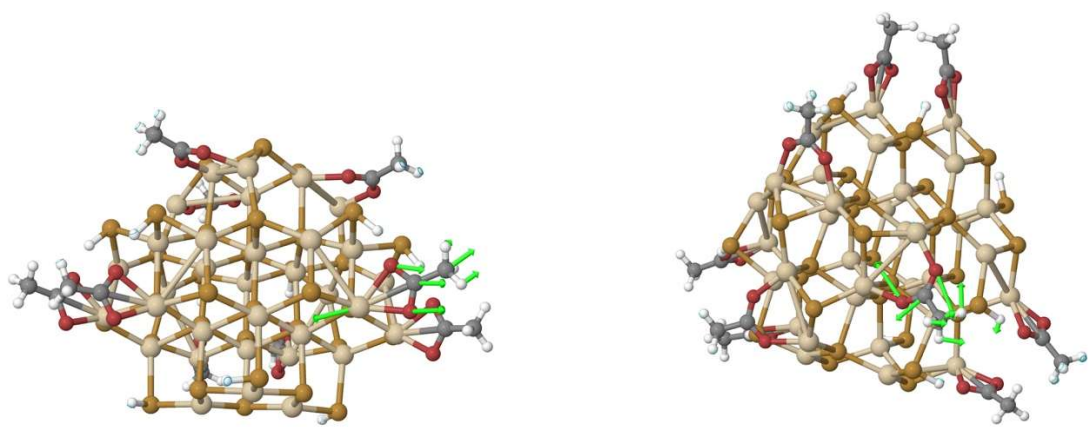

**Figure 4.** Green arrows show two lifting-off vibrational modes of the acetate-passivated QD at 300 cm$^{-1}$ (left) and 497 cm$^{-1}$ (right) associated with relaxation through the cascade of conical intersections (CI$_{X4-X3}$, CI$_{X3-X2}$, and CI$_{X2-X1}$) to the minimum of X1 (Min$_{X1}$).

We now assign the largest peaks in the theoretical spectral density. Figure 3 presents visual representations of the normal modes of the largest peaks in the spectral density, with atoms colored from red (most strongly displaced) to blue (least displaced). The two lower-frequency modes at 93 cm$^{-1}$ and 225 cm$^{-1}$ involve delocalized motions of both core and ligand atoms, while the higher-frequency modes at 1446 cm$^{-1}$ and 1545 cm$^{-1}$ are localized on a small number of specific ligands (see supplementary movie files). The 1446 cm$^{-1}$ mode corresponds to a chelating ligand symmetric stretch, and the 1545 cm$^{-1}$ mode corresponds to a bridging ligand asymmetric stretch. The two intermediate-frequency modes at 300 cm$^{-1}$ and 497 cm$^{-1}$ involve a *lifting-off* motion of one of the bridging ligands. What we refer to here as lifting-off motion are stretching modes of the Cd-O (core-ligand) bonds. This assignment is consistent with the photoinduced ligand detachment and reattachment process reported by Grega, et al. in stearate-passivated CdSe QDs, observed by ultrafast mid-infrared transient absorption spectroscopy (Figure 4 and supplementary movie files). Following these authors, we use the term lifting-off to reflect that fact that, with sufficient excitation, these motions will naturally lead to ligand detachment. Additional modes in this region—631 cm$^{-1}$ and 944 cm$^{-1}$—correspond to $CO_2$ bending motions of specific ligands (Figure 3 and supplementary movie files).

## 3.3. Origin of Core-Ligand Coupling

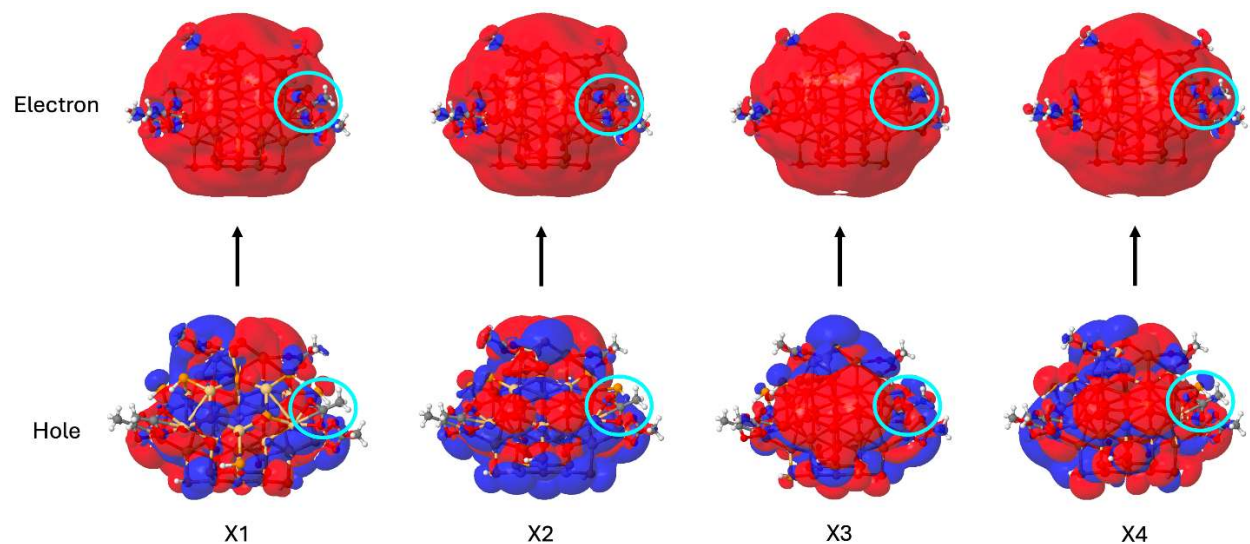

**Figure 5.** Natural transition orbitals representing excitation from the hole to the electron in the four exciton states of the acetate-passivated model QD, computed at the FC geometry. An acetate ligand involved in the 300 cm$^{-1}$ vibrational mode (pictured in Figure 3) is circled in turquoise.

Lastly, we explore the origin of the coupling between core electronic excitation and ligand vibrations that is responsible for the ligand dependence of the cooling rate. In previous work, we identified two possible coupling mechanisms—vibrational and electronic—by which hot carrier cooling can occur in amine-passivated QDs following photoexcitation.[66] Both mechanisms begin with the delocalized electronic excitation of the core atoms. In the vibrational mechanism, the core vibrations are excited via electron-phonon coupling and are in turn harmonically coupled to the ligand atoms. In the electronic mechanism, delocalization of the core electronic excitation onto the ligands directly excites ligand vibrations owing to mixing with the LUMO energy levels.[83] In amine-passivated QDs, we observed evidence of both mechanisms, with the vibrational mechanism dominating at low vibrational frequencies and the electronic mechanism dominating at higher frequencies.

We see similar evidence for the vibrational mechanism at low frequencies in our carboxylate-passivated dots. Figure 3 shows that the vibrations at 93 and 225 cm$^{-1}$ are delocalized across the core and ligands, indicating a strong harmonic coupling. But the higher-frequency modes are much more localized, ruling out the vibrational mechanism.

To understand the mechanism by which energy is coherently transferred to these higher-frequency modes, we examine the natural transition orbitals (NTOs) of the acetate-passivated model shown in Figure 5, computed at the Franck–Condon point for the X1–X4 states. The circled ligand is one that participates in the 300 cm$^{-1}$ mode excited during relaxation. We find that each state exhibits a different degree of delocalization of the excitation onto that ligand. Such delocalization drives ligand vibrational motion in the excited state. Furthermore, differences in the degree of delocalization between states give rise to differences in coupling, resulting in the types of excited-state intersections we describe above.

**Table 1.** Percentage of the integrated spectral density in the low (<300 cm$^{-1}$) and high (>300 cm$^{-1}$) frequency modes in the acetate- and methylamine-passivated model QDs.

|  | Acetate-passivated QD | Methylamine-passivated QD |
| --- | --- | --- |
| Low Frequencies (<300 cm$^{-1}$) | 79% | 35% |
| High Frequencies (>300 cm$^{-1}$) | 21% | 65% |

Having established that both electronic and vibrational mechanisms contribute to energy relaxation in these carboxylate systems, we now quantify the relative importance of each. To this end, Table 1 reports the percentage of the integrated spectral density in the low-frequency (<300 cm$^{-1}$) and high-frequency (>300 cm$^{-1}$) regions for both the acetate-passivated model studied here and the methylamine-passivated QD model studied in our previous work (Cd$_{33}$Se$_{33}$(NH$_2$CH$_3$)$_{21}$;

Table S1).[66] Within our approximations, this is the percentage of energy released into each set of vibrational modes during relaxation. We found that only 21% of the energy is released into high-frequency modes in the acetate-passivated QD, whereas 65% is released into high-frequency modes in the methylamine-passivated QD. This suggests that the electronic mechanism is less pronounced in the carboxylate-passivated QDs compared to the amine-passivated ones. This conclusion is consistent with qualitative trends observed in the experimental results, where greater amplitude is seen in the high-frequency region of the Fourier-transform spectra of amine-passivated QDs[63] than in those of carboxylate-passivated QDs. Additional support for the dominance of the vibrational is visible in the to-be-published 3DES oscillation maps, which show that the LO phonon mixes with the 300-400 cm$^{-1}$ modes that contribute prominently to the excited-state wave packet motion.

## 3.4. Ligand-Truncation Dependence of Model

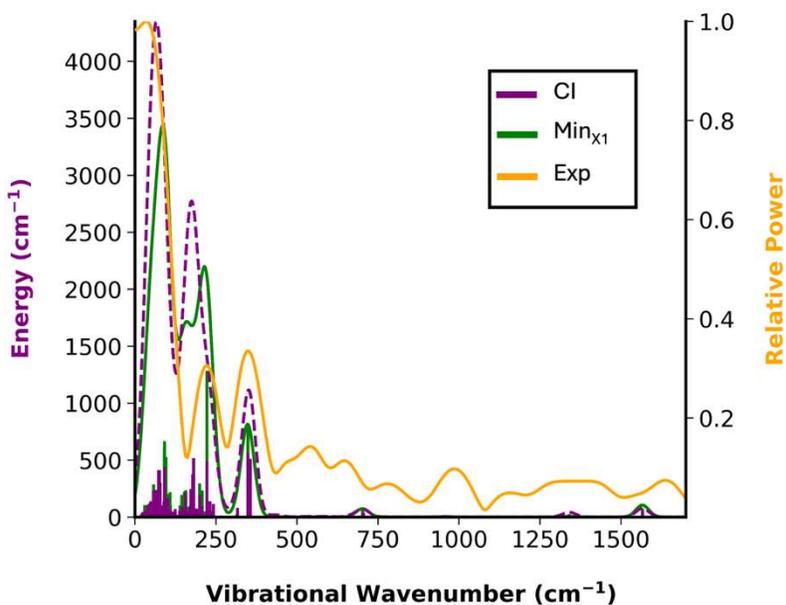

**Figure 6.** Spectral densities (in cm$^{-1}$) of the formate-passivated QD. Purple lines and sticks correspond to the full relaxation path from FC to the minimum of X1 (Min$_{X1}$), passing through CI$_{X4-X3}$, CI$_{X3-X2}$, and CI$_{X2-X1}$. The green line and sticks correspond to a direct linear relaxation pathway from FC to Min$_{X1}$. For comparison, the experimental Fourier transform amplitude spectrum of the X3-PL cross peak, described in the text, is shown in orange.

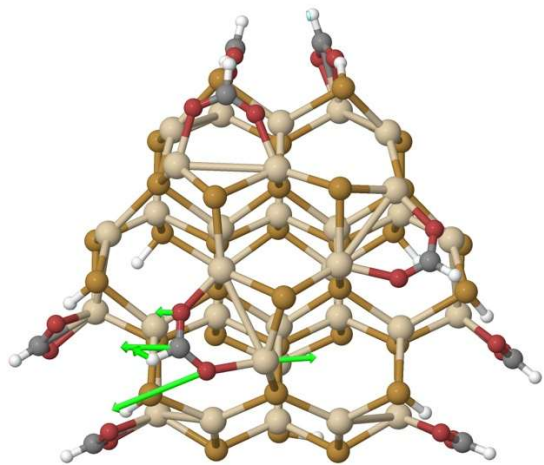

**Figure 7.** Green arrows show a lifting-off vibrational mode at 348 cm$^{-1}$ associated with relaxation of the formate-passivated QD model through the cascade of conical intersections (CI$_{X4-X3}$, CI$_{X3-X2}$ and CI$_{X2-X1}$) to the minimum of X1 (Min$_{X1}$).

Furthermore, we computed the spectral density of the formate-passivated model in the same way (purple line and sticks in Figure 6). The orange line shows the same experimental Fourier-transform spectrum as in Figure 2. As for the acetate system, there is good agreement between the calculated and experimental vibrational modes in the lower-frequency region of both spectra, with the calculated modes at 92 cm$^{-1}$ and 181 cm$^{-1}$ involving delocalized motion over both core and ligand atoms (see supplementary movie files). There is also good agreement between calculated and experimental modes in the higher-frequency region above 1250 cm$^{-1}$, with two main peaks appearing at 1339 cm$^{-1}$ and 1567 cm$^{-1}$. These correspond to a bridging ligand symmetric stretch and a bridging ligand asymmetric stretch, respectively (see supplementary movie files). However, these high-frequency modes are less intense in the calculated spectral density of the formate system than in that of the acetate system. This difference is likely due to the added motion of the methyl group in the acetate ligands compared to the formate ligands. In the intermediate-frequency region, there is a narrow cluster of modes around 348 cm$^{-1}$, which involves a lifting-off motion of one of the bridging ligands, similar to modes observed in the same region of the acetate QD. These lifting-off modes are reminiscent of a photoinduced ligand detachment process reported by Grega, et al[45] (Figure 7 and supplementary movie files), which will be discussed in more detail below.

Though the spectral densities for both models resemble the experimental Fourier transform spectrum, in the formate-passivated case, some modes between 300-1100 cm$^{-1}$ are missing, suggesting that the motion of the atoms beyond the carboxylate group may contribute to hot carrier cooling. In particular, modes observed in the 500-650 cm$^{-1}$ and 900-1000 cm$^{-1}$ region for the acetate-passivated dots are not present in the formate-passivated case. These modes involve motion of the methyl groups of the acetate ligands, therefore their absence is an artifact of the truncation of the ligands.

As mentioned, our calculations suggest the importance of a ligand lifting-off motion in the 300–500 cm$^{-1}$ region in both models, previously observed experimentally by Grega, et al.[45] However, the spectral densities of the two models differ noticeably in this region. To understand the origin of this difference, we further analyzed the modes by categorizing each vibrational motion in this range as one of the following: chelating Cd-O symmetric stretch, chelating Cd-O asymmetric stretch, bridging Cd-O symmetric stretch, or bridging Cd-O asymmetric stretch (Figure 8). The blue dots correspond to vibrational modes in the acetate-passivated model, while the red dots represent those in the formate-passivated model. We found that the symmetric

modes—whether chelating or bridging—are shifted to higher frequencies in the formate system compared to the acetate system by tens of wavenumbers. In contrast, the asymmetric modes are shifted lower in frequency by as much as 200 cm$^{-1}$. These results show that ligand truncation significantly affects the vibrational frequencies in this region.

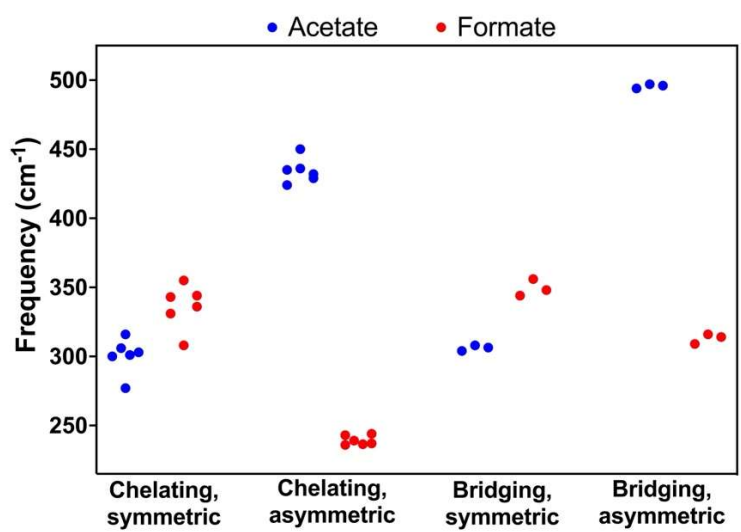

**Figure 8.** Vibrational modes between 300 and 500 cm$^{-1}$ of the acetate- and formate-passivated QD models (in blue and red, respectively), categorized by surface-binding motif and type of vibration: chelating Cd-O symmetric stretch, chelating Cd-O asymmetric stretch, bridging Cd-O symmetric stretch, and bridging Cd-O asymmetric stretch.

### 3.5. Simplified Reaction Coordinate

Up to this point, our calculations have explicitly followed relaxation pathways that pass through all three CIs. However, one might ask how well the calculated spectral density would match the experimental data if we assumed a direct linear path connecting FC to Min$_{X1}$ instead, bypassing the intersections. To investigate this alternative, we computed the spectral densities for direct relaxation to Min$_{X1}$ in both models. The green line and sticks in Figures 2 and 5 represent the spectral densities corresponding to these direct pathways in each model.

In both models, similar spectra are observed whether or not the path passes directly through the conical intersections. In our previous work on hot carrier cooling in amine-passivated QDs, we performed a similar analysis and found that important peaks were missing if the relaxation pathway did not pass through the CIs, indicating that direct relaxation to Min$_{X1}$ did not account for all experimentally observed motion.[66] Thus, in some cases, a relaxation pathway that includes each conical intersection is essential to capture the experimentally observed dynamics. In others, a direct linear path from the FC point to the excited state minimum suffices.

## 3.6. Multiple Relaxation Pathways

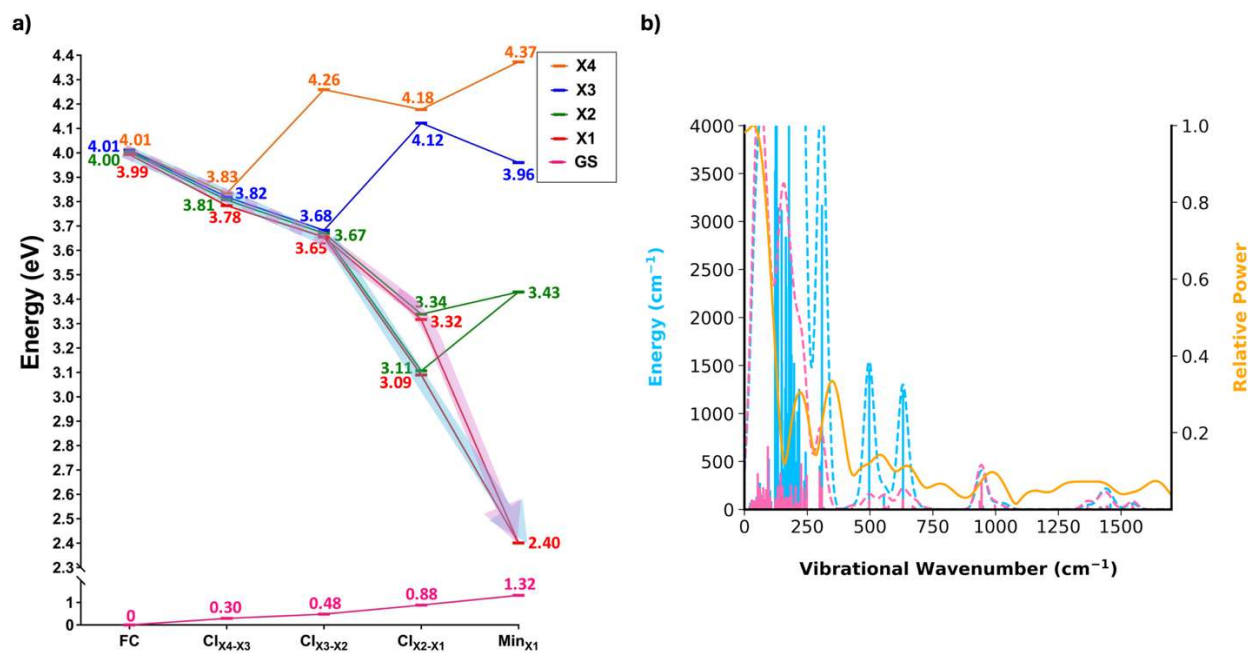

**Figure 9.** Characterization of two relaxation pathways in the acetate-passivated model involving different X2-X1 CIs. In both panels, the original relaxation pathway (described above) is indicated by pink, while the new reaction pathway (replacing CI$_{X2-X1}$ with CI'$_{X2-X1}$) is indicated by blue. a) Energies (in eV) of the ground and first four exciton states at the Franck−Condon (FC), conical intersection (CI$_{X4-X3}$, CI$_{X3-X2}$, and CI$_{X2-X1}$), and X1 minimum (Min$_{X1}$) geometries. Energies are shifted such that the ground state energy at the FC point is zero. b) The spectral densities associated with each pathway (pink and blue) are shown in comparison to the experimental Fourier transform spectrum (orange) described in the text. The y-axis is cut off at 4000 cm$^{−1}$ for clarity.

Given the large number of degrees of freedom, one would expect many possible relaxation pathways to exist, involving various combinations of surface ligands. The pathway presented above is just one example. To explore the possibility of alternative pathways, we identified an additional CI in the acetate-passivated model, CI'$_{X2-X1}$. This alternative CI was located using the Min$_{X1}$ geometry as an initial guess, whereas the original CI$_{X2-X1}$ geometry was optimized using the FC point as the initial guess.

The blue arrow in Figure 9a indicates the newly identified relaxation pathway, while the pink arrow shows the original pathway (from Figure 1b). The corresponding spectral density for the new pathway is shown in Figure 9b (blue sticks and dashed line). Although this alternate pathway passes through a different CI, the peaks in the spectral density appear at essentially the same frequencies as those in the original pathway (pink sticks and dashed lines in Figure 9b). This agreement provides evidence supporting the robustness of our analysis—although it is not feasible to identify every possible relaxation pathway, different chemically similar reaction paths are likely to yield similar spectroscopic signatures. Further work is needed to more rigorously sample the full range of relaxation pathways.

## 4. Conclusion

We investigated the initial ultrafast steps of hot carrier cooling in two model CdSe QDs with different carboxylate surface ligands using a theoretical approach that assumes relaxation occurs via a cascade of conical intersections between pairs of excited states. This model predicts that energy is released into a broad range of vibrational modes during cooling, spanning 0–1600 cm-1. Comparison of these calculated modes with the frequencies of experimentally observed vibronic coherences indicates that our conical intersection-based relaxation model is consistent with experimental data. A simplified model—assuming a linear reaction pathway that bypasses the conical intersections—also produced results in agreement with experiment in this case. However, in our previous work on amine-passivated QDs, the inclusion of conical intersections was essential to accurately reproduce the experimentally observed coherences.[66]

The present study demonstrates that the central result of our prior study—that relaxation through cascades of conical intersections can account for the first ultrafast steps in cooling—can be generalized beyond amine-passivated QDs. Moreover, by comparing the spectral densities of amine- and carboxylate-passivated QDs, we evaluated the influence of different surface ligands on the relaxation mechanism. This comparison suggests that delocalization of the electronic excitation onto the ligands plays a smaller role in driving cooling dynamics in carboxylate-passivated QDs than in the amine-passivated systems of our earlier study.

Additionally, we demonstrated that multiple reaction pathways exist in a single QD by optimizing two distinct CIs connecting X2 to X1 in the acetate-passivated model.  Yet these different pathways predict spectral densities with similar features, suggesting that similar physics is at play.  This similarity also supports the robustness of our approach, even though a vast number of relaxation pathways likely exist in realistic QDs.

## Acknowledgments

This research was supported by grant award DE-SC0021197 from the Solar Photochemistry program of the Office of Basic Energy Sciences, U.S. Department of Energy. We acknowledge Greg Van Patten and Mengliang Zhang for contributions to the corresponding experimental effort. CVH and BGL gratefully acknowledge funding from the Institute for Advanced Computational Science at Stony Brook University. This work used Expanse GPU at the San Diego Supercomputer Center through allocation CHE-140101 from the Advanced Cyberinfrastructure Coordination Ecosystem: Services & Support (ACCESS) program, which is supported by U.S. National Science Foundation grants #2138259, #2138286, #2138307, #2137603, and #2138296.

## Supplementary Materials

A supplementary document presents supplementary figures and a brief methodological description of the 3DES experiments to which we compare in this manuscript.  Additionally, 11 supplementary movie files animating normal modes discussed in the text are included as supplementary data.  The name of each file indicates the wavenumber of the mode presented (e.g. the 92 cm$^{-1}$ mode is presented in *92mode.mov*).

## Author Declaration

The authors declare no competing financial or personal interests.

## Data Availability Statement

The data that support the findings of this study are available from the corresponding author upon reasonable request.